\documentclass[twocolumn,trackchanges]{aastex701}
\graphicspath{{./}{figures/}}
\usepackage{amsmath}  
\usepackage{amsfonts} 
\usepackage{tikz}
\usepackage{multirow} 
\usetikzlibrary{shapes.geometric, arrows}
\usetikzlibrary{positioning}
\begin{document}
\hypersetup{allcolors=black,pdfborder={0 0 0}}

\title{Symbolic Regression Fitting for Roche Lobe Radius}

\author[orcid=0009-0002-7364-7449,sname=Ma,gname=Ye]{Ye Ma}
\affiliation{School of Physics and Astronomy, Sun Yat-sen University, Zhuhai 519082, China}
\email{maye5@mail2.sysu.edu.cn}

\author[orcid=0000-0003-3964-0438,sname=Li,gname=XiaoDong]{Xiao-Dong Li}
\affiliation{School of Physics and Astronomy, Sun Yat-sen University, Zhuhai 519082, China}
\affiliation{CSST Science Center for the Guangdong-HongKong-Macau Great Bay Area, Sun Yat-sen University, Zhuhai 519082, China}
\email{lixiaod25@mail.sysu.edu.cn}

\author[orcid=0000-0002-0378-2023,sname=Ma,gname=Bo]{Bo Ma}
\affiliation{School of Physics and Astronomy, Sun Yat-sen University, Zhuhai 519082, China}
\affiliation{CSST Science Center for the Guangdong-HongKong-Macau Great Bay Area, Sun Yat-sen University, Zhuhai 519082, China}
\email{mabo8@mail.sysu.edu.cn}

\author[orcid=0000-0001-9204-7778,sname=Han,gname=ZhanWen]{ZhanWen Han}
\affiliation{International Centre of Supernovae, Yunnan Key Laboratory, Kunming, 650216, People’s Republic of China}
\affiliation{Yunnan Observatories, Chinese Academy of Sciences, Kunming, 650216, People’s Republic of China}
\email{zhanwenhan@ynao.ac.cn}

\author[orcid=0000-0003-0454-7890,sname=Yu,gname=Cong]{Cong Yu}
\affiliation{School of Physics and Astronomy, Sun Yat-sen University, Zhuhai 519082, China}
\affiliation{CSST Science Center for the Guangdong-HongKong-Macau Great Bay Area, Sun Yat-sen University, Zhuhai 519082, China}
\affiliation{International Centre of Supernovae, Yunnan Key Laboratory, Kunming, 650216, People’s Republic of China}
\affiliation{State Key Laboratory of Lunar and Planetary Sciences, Macau University of Science and Technology, Macau, People’s Republic of China}
\email{yucong@mail.sysu.edu.cn}

\correspondingauthor{Cong Yu}
\email[show]{yucong@mail.sysu.edu.cn}
\begin{abstract}
\textcolor{black}{We present a compact approximation to the dimensionless, volume-equivalent Roche lobe radius $R_{\rm L}/a$, obtained by applying symbolic regression to radii calculated from the full Roche potential. Over $10^{-5}\leq q\leq10^5$, the expression reduces the root mean square error from $1.97\times10^{-3}$ for Eggleton's formula to $5.95\times10^{-4}$, while its maximum relative error is $1.39\%$, compared with Eggleton's $0.83\%$. Calculations at additional mass ratios give a mean relative error of $0.331\%$. The formula is monotonic, recovers the $q^{\frac{1}{3}}$ scaling at small mass ratio, and approaches a large-mass-ratio limit within $0.036\%$ of the limiting Roche model. In a NumPy float64 benchmark of $10^7$ direct formula evaluations, the SR expression takes $0.217$~s compared with $0.450$~s for Eggleton, a speed-up of $2.08$. These results quantify the accuracy and evaluation cost of a compact algebraic alternative for binary-system calculations.}
\end{abstract}

\keywords{Binary systems, Roche lobe, Symbolic regression, Parametric formulas}

\section{Introduction}

\textcolor{black}{In a close binary system, the motion of gas is influenced by the gravity of both stars. In a frame rotating with a circular binary, these gravitational contributions and the centrifugal term define an effective potential. Surfaces on which this potential is constant are called equipotential surfaces.} \textcolor{black}{In the classical Roche model with synchronous rotation,} the Roche lobe of each star is bounded by the critical equipotential surface passing through the inner Lagrangian point, $L_1$, between the stars \citep{1959cbs..book.....K}. If a star expands until its surface reaches this boundary, gas from its outer layers can flow through $L_1$ toward the companion, transferring mass between the stars. \textcolor{black}{This process is known as Roche-lobe overflow} and can alter the subsequent evolution of the binary.

\textcolor{black}{The Roche lobe is not spherical, so its size is described by the volume-equivalent radius $R_{\rm L}$: the radius of a sphere enclosing the same volume as the lobe} \citep{1983ApJ...268..368E}. \textcolor{black}{For the lobe surrounding a star of mass $M_1$ with a companion of mass $M_2$, the ratio $R_{\rm L}/a$ depends only on the mass ratio $q=M_1/M_2$ within this model, where $a$ is the orbital separation.} A reliable and inexpensive approximation to this ratio is useful in binary evolution calculations, population synthesis, and hydrodynamic modeling. Its relevance remains clear in recent studies of interacting binaries, stripped stars, and the onset of mass-transfer instabilities \citep{2024ApJ...969....8S,2024A&A...683A..37Y,2025A&A...703A.121T}.

The approximation of \citet{1983ApJ...268..368E} is used because its relative error is below about $1\%$ over the full mass-ratio range, $0<q<\infty$. Its fractional powers and logarithm motivate testing compact algebraic alternatives in applications requiring many radius evaluations. Section~\ref{sec:timing} compares their computational costs in NumPy batch calculations.

\textcolor{black}{Symbolic regression searches for both the structure of a mathematical expression and its numerical constants. This differs from conventional parameter fitting, which adjusts constants within a prescribed functional form.} Its explicit expressions can be inspected, simplified, and implemented, but their accuracy and physical interpretation require validation. \textcolor{black}{Reviews discuss the diversity of search strategies and the challenges of complexity control, robustness, and interpretation} \citep{MakkeChawla2024,2026RSPTA.38420240600B}. \textcolor{black}{An influential physics-inspired approach, AI Feynman, combines neural-network approximations with tests for simplifying properties such as symmetry and separability} \citep{UdrescuTegmark2020}. \textcolor{black}{PySR provides a flexible evolutionary search over expression trees, interleaving evolution with simplification and numerical optimization of constants} \citep{2023arXiv230501582C}.

\textcolor{black}{Recent methodological work uses learned representations and scientific prior knowledge to guide the search. LLM-SR represents candidate equations as programs and combines language-model proposals with evolutionary search} \citep{Shojaee2025LLMSR}, \textcolor{black}{whereas LLM-Feynman combines domain knowledge with feature construction and guided equation search} \citep{Song2025LLMFeynman}. \textcolor{black}{Neural approaches using order-invariant embeddings and sparse decoding illustrate another direction, while highlighting that good numerical predictions do not necessarily imply exact recovery of the underlying expression} \citep{Malik2025}. \textcolor{black}{Broader discussions of AI-assisted scientific discovery distinguish predictive utility, evaluation, and explanatory knowledge} \citep{ReddyShojaee2025,CruzAguilar2026}. We apply PySR's evolutionary search to approximate the classical Roche-model relation.

In this paper, we use symbolic regression to derive a compact approximation for the Roche lobe radius. \textcolor{black}{We construct the reference dataset by locating the inner Lagrangian point, solving for the critical equipotential surface of the full Roche potential, and evaluating the enclosed volume with adaptive numerical quadrature (Section~\ref{subsec:roche}). The resulting volume-equivalent radii, normalized by the orbital separation, are calculated independently of the Eggleton approximation and provide the targets for symbolic regression.} We search for simple expressions with a restricted operator set and compare the resulting formula with the Eggleton form and coefficient-optimized variants of both families. We assess each formula with absolute and relative error metrics, together with timing benchmarks.

This paper is structured as follows. Section~2 reviews the numerical method for generating benchmark data and the symbolic regression technique. Section~3 presents the discovered formula and compares it with existing approximations. Section~4 summarizes the findings.

\section{Methodology} \label{sec:style}
\subsection{Roche Lobe Radius Calculation} \label{subsec:roche}

Following \citet{1959cbs..book.....K}, two point masses $M_1$ and $M_2$ in a circular orbit are placed at $z = 0$ and $z = 1$ on the line of
centers, with mass ratio $Q = M_2/M_1=1/q$. \textcolor{black}{All spatial coordinates in this calculation are expressed in units of the orbital separation $a$, so the numerical volume $V$ is the physical lobe volume divided by $a^3$.} The Roche potential in a co-rotating frame is given by Equation~\ref{eq:potential}. The equivalent Roche lobe radius is obtained through the numerical procedure below.
Calculate the Lagrange point $z_0$ by solving the equation:

\begin{equation} \label{eq:lagrange_eq}
(1+Q)z_0^5 - (2+3Q)z_0^4 + (1+3Q)z_0^3 - z_0^2 + 2z_0 - 1 = 0
\end{equation}

Compute the critical Roche potential $\Omega_0$ at the inner Lagrangian point:

\begin{equation} \label{eq:omega0}
    \Omega_0 = \frac{1}{z_0} + \frac{Q}{1-z_0} + \frac{1+Q}{2}z_0^2 - Qz_0
\end{equation}

Locate another point $z_1$ on the $z$-axis where the potential equals $\Omega_0$ by solving:

\begin{equation} \label{eq:z1}
    \Omega(0, z, 0) = \Omega_0
\end{equation}

where $z_1$ is the outer intersection of the same equipotential surface
with the $z$ axis on the side of $M_1$ opposite to $M_2$.

For each $z \in [z_1, z_0]$ and $\theta \in [0, 2\pi]$, determine $R(\theta, z)$ by solving:
\begin{equation} \label{eq:R_theta_z}
    \Omega(R, z, \theta) = \Omega_0
\end{equation}

where the potential function $\Omega(R, z, \theta)$ is defined as:

\begin{equation} \label{eq:potential}
    \begin{split}
        \Omega(R, z, \theta) ={}& \frac{1}{\sqrt{R^2 + z^2}} + \frac{Q}{\sqrt{R^2 + (1-z)^2}} \\
        &+ \frac{1+Q}{2}R^2\sin^2\theta + \frac{1+Q}{2}z^2 - Qz
    \end{split}
\end{equation}

Calculate the volume $V$ of the Roche lobe:
\begin{equation} \label{eq:volume_integral}
    V = \int_{z_1}^{z_0} \int_0^{2\pi} \frac{1}{2}R^2(\theta, z)  d\theta dz
\end{equation}

Determine the dimensionless Roche lobe radius $R_{\rm L}/a$ using the spherical volume formula:
\begin{equation} \label{eq:equivalent_radius}
    V = \frac{4}{3}\pi (R_{\rm L}/a)^3 \quad \Rightarrow \quad R_{\rm L}/a = \left( \frac{3V}{4\pi} \right)^{\frac{1}{3}}
\end{equation}

\textcolor{black}{The nonlinear equations are solved using SciPy's \texttt{newton} routine without analytic derivatives, which uses the secant method. We retain its default absolute step tolerance of $1.48\times10^{-8}$, zero relative tolerance, and maximum of 50 iterations. The initial guesses are 0.5 for $z_0$ and $-z_0$ for $z_1$. At each $z$, we first solve for $R(0,z)$ from an initial guess of 0.5 and then use this solution to initialize the surface-radius calculation at other angles. The volume integral is evaluated with SciPy's \texttt{dblquad} routine using adaptive quadrature and its default absolute and relative tolerance parameters, both $1.49\times10^{-8}$} \citep{2020NatMe..17..261V}. \textcolor{black}{These parameters control the numerical routines; they are not independent estimates of the uncertainty in the final radius.}

\textcolor{black}{The dataset supplied to symbolic regression consists of pairs $(q,R_{\rm L}/a)$ covering approximately $10^{-5}\leq q\leq 3.69\times10^5$. The mass ratios are distributed non-uniformly in logarithmic space, with denser sampling between $q=1$ and $100$. All saved pairs are used in the symbolic-regression search. The comparisons against these numerical data measure approximation errors on the reference samples, rather than performance on a separate test set.}

\textcolor{black}{The reference arrays, a standalone implementation of the numerical generation method, the symbolic-regression search script, and the independent validation programs are available in a public repository.}\footnote{\url{https://github.com/Iceberg0716/SRLochelobe}}

\begin{figure}[!ht]
  \centering
  \includegraphics[width=0.95\columnwidth]{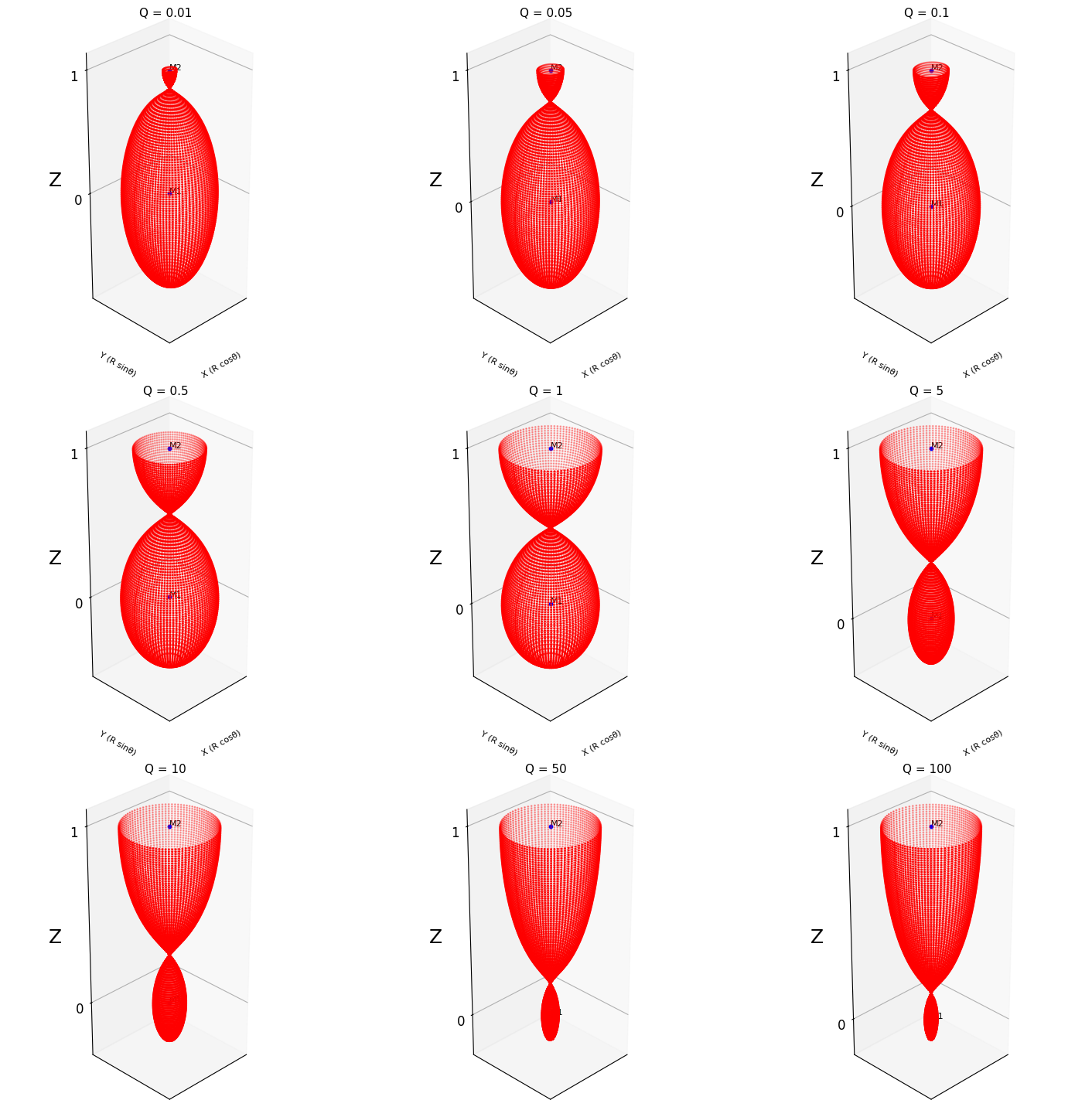}
  \caption{Three-dimensional equipotential contours for binary systems with different mass ratios ($Q = M_2/M_1$). Each subplot shows the gravitational potential surface at the inner Lagrange point energy level, with the primary mass ($M_1$) at $z=0$, secondary mass ($M_2$) at $z=1$, and the inner Lagrange point ($z_0$) and outer axial intersection ($z_1$) marked. The variation in contour shape demonstrates how the mass ratio affects the gravitational potential geometry, from extreme mass ratios ($Q = 0.01, 100$) to equal masses ($Q = 1$).}
  \label{fig:Loche3d} 
\end{figure}

Figure~\ref{fig:Loche3d} shows the resulting equipotential contours computed by this procedure for representative mass ratios spanning $Q = 0.01$ to $Q = 100$.

\subsection{Symbolic Regression Technique} \label{subsec:sr}

\textcolor{black}{For each sample $j=1,\ldots,N$, the sole input is the dimensionless mass ratio $q_j=M_1/M_2$ and the target is $y_j=(R_{\rm L}/a)_j$, where $N$ denotes the number of training samples. The dimensional masses $M_1,M_2$ and separation $a$ are not supplied as separate input features. Dimensional consistency is inherited from the normalized data representation, rather than discovered by the search from dimensional inputs. The target $R_{\rm L}/a$ is distinct from the local cylindrical surface coordinate $R(\theta,z)$ used in the volume integral.}

\subsubsection{Genetic Programming Framework}

This study uses symbolic regression based on genetic programming (GP) to discover analytical expressions from data. Candidate expressions are encoded as expression trees: leaf nodes represent the input variable $q$ or numerical constants, and internal nodes represent mathematical operators drawn from a predefined set.

The operator set determines the space of discoverable expressions. To search for a simple Roche lobe radius formula similar in form to the Eggleton relation, we constrain the operators to the four basic arithmetic operations and two roots:
\begin{equation}
    \mathcal{F}_{\text{bin}} = \{+, -, \times, \div\}, \quad
    \mathcal{F}_{\text{unary}} = \{\sqrt{\cdot}, \sqrt[3]{\cdot}\}.
\end{equation}
This operator set restricts the search to algebraic expressions. We set the maximum expression complexity to \texttt{maxsize}=20 to favor compact formulas. \textcolor{black}{Complexity control does not by itself establish predictive accuracy; the resulting expression must be checked against numerical reference values and physical limits.}

\subsubsection{Evolutionary Search Process}

\begin{figure}[!htbp]
    \centering
    \resizebox{\columnwidth}{!}{%
    \begin{tikzpicture}[
        node distance=0.6cm,   
        startstop/.style={ellipse, minimum width=2.6cm, minimum height=0.75cm,
                          text centered, draw=black, fill=gray!10},
        process/.style={rectangle, minimum width=4.1cm, minimum height=0.95cm,
                        text centered, draw=black, text width=4.2cm, align=center, fill=white},
        decision/.style={diamond, minimum width=3.0cm, minimum height=0.95cm,
                         text centered, draw=black, text width=2.8cm, align=center, fill=white},
        arrow/.style={thick,->,>=stealth}
    ]

    \node (start) [startstop] {Start};
    \node (init)  [process, below=of start] {Randomly generate initial\\
        expression population\\[2pt]
        \footnotesize \textit{ops:} $+$, $-$, $\times$, $\div$, $\sqrt{\ }$, $\sqrt[3]{\ }$};
    \node (eval)  [process, below=of init] {Evaluate fitness of each expression\\
        (e.g., MSE)};
    \node (dec)   [decision, below=of eval, yshift=-0.15cm] {Termination condition\\
        satisfied?};
    \node (best)  [process, below=of dec, yshift=-0.15cm] {Selected expression:\\[2pt]
        $\displaystyle R_{\rm SR} = \left(\frac{q}{7.6\sqrt{q} + 1.85q + 8.8}\right)^{\frac{1}{3}}$};
    \node (stop)  [startstop, below=of best, yshift=-0.1cm] {End};

    \node (select) [process, right=of dec, xshift=0.6cm] {Selection\\
        (tournament)};
    \node (cross)  [process, above=1.1cm of select] {Crossover\\
        (swap subtrees)};
    \node (mut)    [process, above=1.1cm of cross] {Mutation\\
        (randomly modify node)};
    \node (newgen) [process, above=1.1cm of mut] {Generate new population};

    \draw [arrow] (start) -- (init);
    \draw [arrow] (init)  -- (eval);
    \draw [arrow] (eval)  -- (dec);
    \draw [arrow] (dec)   -- node[anchor=east, xshift=-0.2cm] {Yes} (best);
    \draw [arrow] (best)  -- (stop);
    \draw [arrow] (dec)   -- node[anchor=south, yshift=0.1cm] {No} (select);
    \draw [arrow] (select) -- (cross);
    \draw [arrow] (cross)  -- (mut);
    \draw [arrow] (mut)    -- (newgen);
    \draw [arrow] (newgen.west) -- ++(-0.55,0) |- (eval.east);

    \end{tikzpicture}
    }
    \caption{Schematic evolutionary search for symbolic regression with PySR; simplification and constant refinement are summarized in the population update.
             Operator set: $+,-,\times,\div,\sqrt{\ },\sqrt[3]{\ }$; Loss: MSE;
              Search iterations: 500; Selected formula: $R_{\rm SR} = (\frac{q}{7.6\sqrt{q}+1.85q+8.8})^{\frac{1}{3}}$.}
    \label{fig:pysr_gp_flow}
\end{figure}
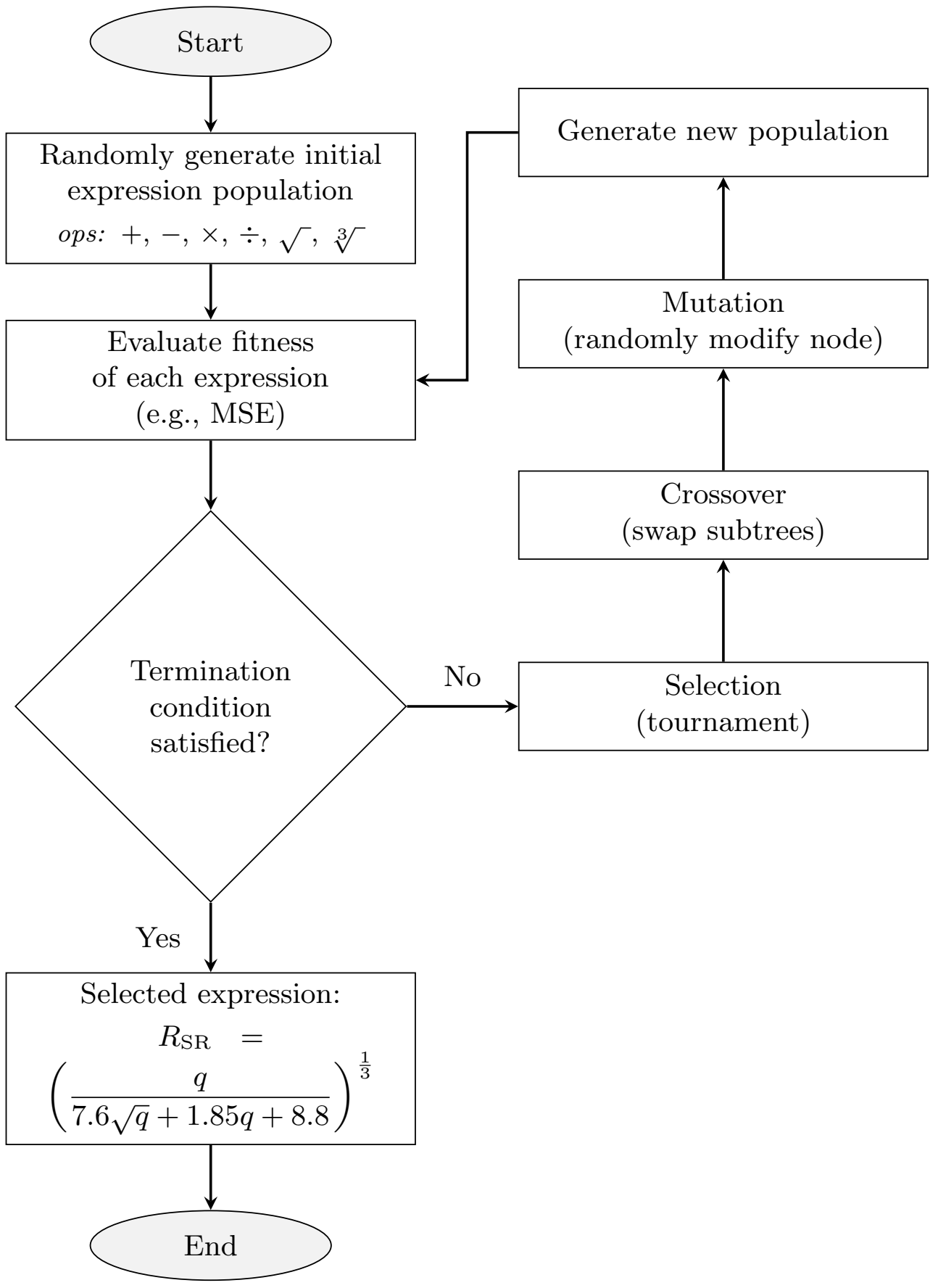

The open-source \texttt{PySR} implementation evolves populations of expression trees using variation, selection, \textcolor{black}{simplification, and optimization of numerical constants} \citep{2023arXiv230501582C}. Figure~\ref{fig:pysr_gp_flow} summarizes the search. \textcolor{black}{A candidate $f_i$ maps $q$ to a predicted normalized radius; $i$ labels the candidate expression and $j$ labels the numerical data sample.} We specify an elementwise squared-error loss, giving the mean squared error
\begin{equation}
    \mathrm{MSE}(f_i) = \frac{1}{N} \sum_{j=1}^{N} \left[f_i(q_j) - y_j\right]^2.
\end{equation}
All samples enter this loss with equal weight, so densely sampled mass-ratio regions contribute more strongly to the fit.

Tournament selection favors competitive candidates, while mutation and crossover alter operators, constants, or subexpressions. \textcolor{black}{PySR simplifies expressions and refines their constants.} \textcolor{black}{We run \texttt{niterations}=500 search iterations; an iteration is not a single genetic-programming generation.} The search retains candidate expressions at different complexities, allowing accuracy and compactness to be compared. \textcolor{black}{The supplied script leaves the remaining search and model-selection settings at the library defaults and exports the selected expression with \texttt{model.sympy()}. We do not impose an additional numerical acceptance threshold in that script.}

\section{Results} \label{sec:floats}

\textcolor{black}{Throughout this section, $R_{\rm SR}(q)$, $R_{\rm Egg}(q)$, and their optimized variants denote dimensionless approximations to $R_{\rm L}/a$. Multiplication by the orbital separation $a$ recovers a physical radius. Their absolute-error statistics are dimensionless, and their relative errors are reported as percentages.}

Using symbolic regression on the numerical dataset, we obtained the following empirical approximation, \textcolor{black}{evaluated over $10^{-5}\leq q=M_1/M_2\leq10^5$}:

\begin{equation}
R_{\rm SR}(q) = \left(\frac{q}{7.6\sqrt{q} + 1.85q + 8.8}\right)^{\frac{1}{3}}
\end{equation}

\begin{figure*}[t]
\centering
\includegraphics[width=\textwidth]{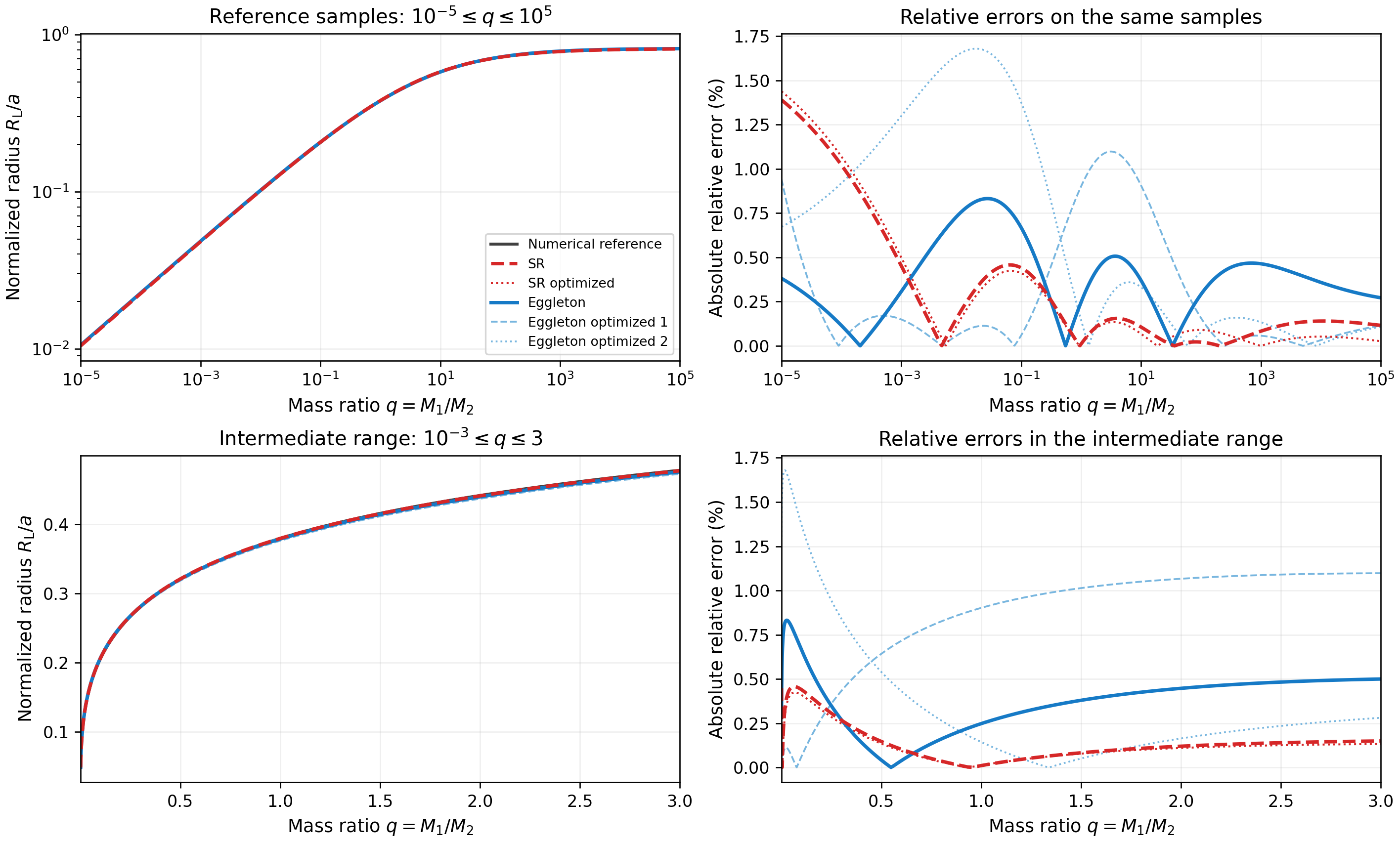}
\caption{
Comparison of dimensionless Roche lobe radius approximations against the numerical reference. The left panels show $R_{\rm L}/a$ and the right panels show absolute relative errors in percent. The upper panels use the samples with $10^{-5}\leq q\leq10^5$, and the lower panels show their subset with $10^{-3}\leq q\leq3$. The SR and original Eggleton formulas are the thick red dashed and blue solid curves, respectively. Their coefficient-adjusted variants use thinner dotted or dashed lines in the same color families; the numerical reference is dark gray.
\label{fig:general}}
\end{figure*}

We compare this expression with the dimensionless Eggleton approximation \citep{1983ApJ...268..368E}:
\begin{equation}
    R_{\rm Egg} = \frac{0.49 q^{\frac{2}{3}}}{0.6 q^{\frac{2}{3}} + \ln(1 + q^{\frac{1}{3}})} \quad \text{(Eggleton 1983)}
\end{equation}
The SR expression replaces the logarithmic denominator term with square-root and linear terms. \textcolor{black}{The evaluation uses the displayed coefficients for all formulas and the reference samples within $10^{-5}\leq q\leq10^5$. Mean errors are arithmetic averages over these samples.} Figure~\ref{fig:general} shows the resulting curves and residuals. The comparison focuses on the SR and Eggleton functional forms and their coefficient-adjusted variants; the older piecewise approximation of \citet{1971ARA&A...9..183P} is omitted.

To refine the analytical expressions, we conducted parameter optimization for both the symbolic regression-derived formula and the classical Eggleton formulation. Using the Levenberg-Marquardt algorithm implemented via SciPy's \texttt{curve\_fit} function \citep{2020NatMe..17..261V}, we performed nonlinear least-squares minimization against our numerical benchmark dataset. \textcolor{black}{The SR functional family is $R_{\rm SR}(q;A,B,C)=[q/(A\sqrt{q}+Bq+C)]^{\frac{1}{3}}$, where $A$, $B$, and $C$ are dimensionless adjustable coefficients. Uppercase letters distinguish these coefficients from the orbital separation $a$.} The coefficient-adjusted SR expression used in the comparisons is
\begin{equation}
R_{\rm SR,opt}(q) = \left(\frac{q}{7.593\sqrt{q} + 1.845q + 8.813}\right)^{\frac{1}{3}}
\end{equation}

We applied the same optimization technique to the Eggleton formula, refining its numerical coefficients to provide a fair baseline for assessing how the symbolic-regression functional form compares in accuracy. This produced two optimized variants:
\begingroup
\begin{align*}
R_{\rm Egg,opt1}(q) &= \frac{0.48547\,q^{\frac{2}{3}}}{0.596745\,q^{\frac{2}{3}} + \ln(0.999511 + q^{\frac{1}{3}})}, \\
R_{\rm Egg,opt2}(q) &= \frac{0.495277\,q^{\frac{2}{3}}}{0.6088\,q^{\frac{2}{3}} + \ln(1 + q^{\frac{1}{3}})},
\end{align*}
\endgroup
Tables~\ref{tab:errors} and \ref{tab:errors_typical} compare the five formulas over the main and restricted mass-ratio intervals, with metric definitions included in each table.

\begingroup
\begin{table}[htbp]
\centering
\setlength{\tabcolsep}{3pt}
\caption{Errors for $10^{-5}\leq q\leq10^5$. \textcolor{black}{For each sample $i$, $y_i$ is the numerical reference for $R_{\rm L}/a$ and $\hat y_i$ is the predicted value. $N$ denotes the number of samples in this interval. Absolute errors are dimensionless; the relative errors defined below are displayed as percentages.} Underlines identify minima before rounding.\label{tab:errors}}
\resizebox{\columnwidth}{!}{%
\begin{tabular}{lccccc}
\hline
Formula & MaxAE & RMSE & MAE & MeanRel & MaxRel \\
{\scriptsize} & $\scriptstyle \max|\hat y_i-y_i|$ & $\scriptstyle \sqrt{\frac{1}{N}\sum(\hat y_i-y_i)^2}$ & $\scriptstyle \frac{1}{N}\sum|\hat y_i-y_i|$ & $\scriptstyle \frac{1}{N}\sum\left|\frac{\hat y_i-y_i}{y_i}\right|$ & $\scriptstyle \max\left|\frac{\hat y_i-y_i}{y_i}\right|$ \\[4pt]
\hline
$R_{\rm SR}$ & $1.13\times10^{-3}$ & \textcolor{black}{$5.95\times10^{-4}$} & \textcolor{black}{$4.80\times10^{-4}$} & \textcolor{black}{$0.2556\%$} & $1.39\%$ \\
$R_{\rm SR,opt}$ & \underline{$8.60\times10^{-4}$} & \textcolor{black}{\underline{$4.24\times10^{-4}$}} & \textcolor{black}{\underline{$3.69\times10^{-4}$}} & \textcolor{black}{\underline{$0.2463\%$}} & $1.44\%$ \\
$R_{\rm Egg}$ & $3.63\times10^{-3}$ & $1.97\times10^{-3}$ & \textcolor{black}{$1.59\times10^{-3}$} & $0.3600\%$ & \underline{$0.83\%$} \\
$R_{\rm Egg,opt1}$ & $5.58\times10^{-3}$ & \textcolor{black}{$2.68\times10^{-3}$} & \textcolor{black}{$1.75\times10^{-3}$} & \textcolor{black}{$0.3790\%$} & $1.10\%$ \\
$R_{\rm Egg,opt2}$ & $2.83\times10^{-3}$ & \textcolor{black}{$1.25\times10^{-3}$} & \textcolor{black}{$1.00\times10^{-3}$} & \textcolor{black}{$0.5079\%$} & $1.68\%$ \\
\hline
\end{tabular}
}
\end{table}
\endgroup

\begingroup
\begin{table}[htbp]
\centering
\setlength{\tabcolsep}{3pt}
\caption{Errors for $10^{-3}\leq q\leq10^3$. \textcolor{black}{For each sample $i$, $y_i$ is the numerical reference for $R_{\rm L}/a$ and $\hat y_i$ is the predicted value. $N$ denotes the number of samples in this interval. Absolute errors are dimensionless; the relative errors defined below are displayed as percentages.} Underlines identify minima before rounding.\label{tab:errors_typical}}
\resizebox{\columnwidth}{!}{%
\begin{tabular}{lccccc}
\hline
Formula & MaxAE & RMSE & MAE & MeanRel & MaxRel \\
{\scriptsize} & $\scriptstyle \max|\hat y_i-y_i|$ & $\scriptstyle \sqrt{\frac{1}{N}\sum(\hat y_i-y_i)^2}$ & $\scriptstyle \frac{1}{N}\sum|\hat y_i-y_i|$ & $\scriptstyle \frac{1}{N}\sum\left|\frac{\hat y_i-y_i}{y_i}\right|$ & $\scriptstyle \max\left|\frac{\hat y_i-y_i}{y_i}\right|$ \\[4pt]
\hline
$R_{\rm SR}$ & $9.21\times10^{-4}$ & $4.82\times10^{-4}$ & \underline{$3.91\times10^{-4}$} & $0.1305\%$ & \underline{$0.46\%$} \\
$R_{\rm SR,opt}$ & \underline{$8.60\times10^{-4}$} & \underline{$4.75\times10^{-4}$} & $4.15\times10^{-4}$ & \underline{$0.1300\%$} & $0.50\%$ \\
$R_{\rm Egg}$ & $3.63\times10^{-3}$ & $1.85\times10^{-3}$ & $1.57\times10^{-3}$ & $0.3960\%$ & $0.83\%$ \\
$R_{\rm Egg,opt1}$ & $5.58\times10^{-3}$ & $3.26\times10^{-3}$ & $2.49\times10^{-3}$ & $0.4956\%$ & $1.10\%$ \\
$R_{\rm Egg,opt2}$ & $2.83\times10^{-3}$ & $1.49\times10^{-3}$ & $1.30\times10^{-3}$ & $0.5342\%$ & $1.68\%$ \\
\hline
\end{tabular}
}
\end{table}
\endgroup

\subsection{\textcolor{black}{Limiting Behavior and Independent Validation}}

\textcolor{black}{For all $q>0$, the denominator is positive, and differentiating $R_{\rm SR}^3$ gives a positive derivative proportional to $3.8\sqrt{q}+8.8$. The normalized radius increases monotonically with mass ratio. Its limiting behavior is}
\begin{equation}
\color{black}
\begin{aligned}
R_{\rm SR}(q)&\sim 0.484365q^{\frac{1}{3}}, && q\to0,\\
R_{\rm SR}(q)&\to 0.814597, && q\to\infty.
\end{aligned}
\label{eq:sr_limits}
\end{equation}
\textcolor{black}{The small-$q$ exponent is consistent with balancing the gravity of the lower-mass star against the companion's tidal field, which gives a characteristic size proportional to $q^{\frac{1}{3}}$. The fitted prefactor is $0.484365$, and the square-root term controls the transition between the two limits.}

\textcolor{black}{Taking $Q=1/q\to0$ in Equation~\ref{eq:potential} gives a limiting critical potential of $\frac{3}{2}$. Let $s$ be the spherical radius in units of $a$ and let $\mu$ be the cosine of the polar angle measured from the rotation axis. The connected inner surface satisfies}
\begin{equation}
\color{black}
\frac{1}{s}+\frac{1-\mu^2}{2}s^2=\frac{3}{2},
\qquad \frac{2}{3}\leq s(\mu)\leq1.
\label{eq:limiting_surface}
\end{equation}
\textcolor{black}{Axisymmetry gives the limiting volume-equivalent radius}
\begin{equation}
\color{black}
\lim_{q\to\infty}R_{\rm L}/a
=\left[\int_0^1s(\mu)^3\,d\mu\right]^{\frac{1}{3}}
\simeq0.814886.
\label{eq:limiting_radius}
\end{equation}
\textcolor{black}{The SR limit differs from this numerical value by approximately $0.0354\%$, compared with $0.2186\%$ for Eggleton's limiting value $0.816667$.}

\textcolor{black}{To test predictions between the training samples, we calculated new reference radii at logarithmically spaced mass ratios over $10^{-4.973}\leq q\leq10^{4.967}$, none coincident with an original sample. The full Roche potential was solved using bracketed root finding and tensor-product Gauss--Legendre quadrature. Symmetry reduces the angular integration to $[0,\pi/2]$; increasing the number of quadrature points in each direction from 48 to 96 changed the radii by at most $9.45\times10^{-8}$ in relative terms. SR, optimized SR, and Eggleton have mean relative errors of $0.331\%$, $0.321\%$, and $0.376\%$, and maximum relative errors of $1.38\%$, $1.43\%$, and $0.83\%$, respectively.}

\subsection{Evaluation Cost}\label{sec:timing}

\textcolor{black}{We benchmarked direct, vectorized NumPy~2.2.6 implementations in Python~3.13.7 on Windows~11 with an Intel Core i7-13650HX processor. Each call processed the same contiguous array of $10^7$ float64 mass ratios, with $\log_{10}q$ sampled uniformly on $[-5,5]$ using PCG64 and seed~20261016. Both implementations use the elementwise power operator for fractional exponents. Eggleton's expression evaluates $q^{\frac{2}{3}}$ separately in the numerator and denominator and uses \texttt{log} for $\ln(1+q^{\frac{1}{3}})$. The SR expression uses \texttt{sqrt} for $\sqrt{q}$ and a final power of $\frac{1}{3}$.}

\textcolor{black}{Each formula received two full-size warm-up calls followed by ten timed calls, with the evaluation order rotated across benchmark rounds. We measured formula evaluation and array allocation with \texttt{perf\_counter\_ns}, excluding input generation, warm-up, validation, and file output. Table~\ref{tab:benchmark} reports median times and interquartile ranges; speed-up is the ratio of the Eggleton and SR medians. The benchmark script, \texttt{benchmark\_implementations.py}, and per-trial records accompany the public code release.}

\begingroup
\begin{table}[htbp]
\centering
\setlength{\tabcolsep}{3pt}
\caption{\textcolor{black}{Wall-clock times for direct NumPy batch evaluation of $10^7$ float64 inputs. Medians and interquartile ranges (IQR; 25th--75th percentiles) are from ten trials per formula.}\label{tab:benchmark}}
\resizebox{\columnwidth}{!}{%
\begin{tabular}{lccc}
\hline
Formula & Median (s) & IQR (s) & Speed-up \\
\hline
$R_{\rm Egg}$ & \textcolor{black}{0.450} & \textcolor{black}{0.448--0.452} & $1.00\times$ \\
$R_{\rm SR}$ & \textcolor{black}{0.217} & \textcolor{black}{0.214--0.221} & \textcolor{black}{$2.08\times$} \\
\hline
\end{tabular}
}
\end{table}
\endgroup

\textcolor{black}{The error comparisons, limiting behavior, and timing measurements characterize a compact empirical surrogate for the classical Roche model. They do not establish a new physical law, a mass-transfer mechanism, or a first-principles derivation of the fitted coefficients. Accuracy outside the validated mass-ratio interval remains to be assessed.}

\section{Conclusion}

We have presented an empirical approximation to the dimensionless Roche lobe radius obtained through symbolic regression,
\[
R_{\rm SR} = \left(\frac{q}{7.6\sqrt{q}+1.85q+8.8}\right)^{\frac{1}{3}}.
\]
The SR expressions reduce absolute errors relative to Eggleton over $10^{-5}\leq q\leq10^5$, while Eggleton retains the smaller maximum relative error. \textcolor{black}{Over $10^{-3}\leq q\leq10^3$, the original SR expression also improves the maximum relative error.}

\textcolor{black}{The expression is positive and monotonic, reproduces the small-mass-ratio $q^{\frac{1}{3}}$ scaling, and approaches the limiting Roche-model radius to within $0.0354\%$. The independent validation gives mean relative errors of $0.331\%$ for SR and $0.376\%$ for Eggleton.}

\textcolor{black}{In the NumPy batch benchmark of direct formula evaluation, SR achieves a $2.08$-fold speed-up over Eggleton.} Its compact algebraic form provides a practical option for binary-system calculations requiring repeated radius evaluations.



\begin{acknowledgments}

This work is supported by the National Key R\&D Program of China (grant No.2024YFA1611803), the National Natural Science Foundation of China (grant No.12403071,12373071), and the science research grants from the China Manned Space Project(CMS-CSST-2025-A16, CMS-CSST-2021-B09, CMS-CSST-2021-B12, and CMS-CSST-2021-A10).

C.Y. has been supported by the National Natural Science Foundation of China (grants 11521303, 11733010, 11873103, and 12373071).

We acknowledge the Beijing Super Cloud Center (BSCC) and Beijing Beilong Super Cloud Computing Co., Ltd (http://www.blsc.cn/) for providing HPC resources that have contributed to the research results presented in this paper.

\end{acknowledgments}

\bibliographystyle{aasjournal}
\bibliography{references/sample701}

\end{document}